\documentclass[aps,twocolumn,groupedaddress]{revtex4}
\usepackage{psfig}
\begin{document}
\bibliographystyle{apsrev}

\preprint{Biochem 01/42}

\title{Finite-sample frequency distributions originating from an equiprobability distribution}

\author{Thorsten P\"oschel}
\email[]{thorsten.poeschel@charite.de}
\homepage[]{summa.physik.hu-berlin.de/~kies}
\affiliation{Humboldt-Universit\"at zu Berlin, Charit\'e,
  Institut f\"ur Biochemie, Monbijoustra{\ss}e 2, D-10117 Berlin,
  Germany}
\author{Jan A. Freund}
\email[]{freund@physik.hu-berlin.de}
\homepage[]{summa.physik.hu-berlin.de/~janf}
\affiliation{Humboldt-Universit\"at zu Berlin, Institut f\"ur Physik, Invalidenstra\ss e 110, D-10115 Berlin, Germany}

\date{\today}

\begin{abstract}
Given an equidistribution for probabilities $p(i)=1/N$, 
$i=1\dots N$.  What is the expected corresponding rank ordered frequency distribution
$f(i)$, $i=1\dots N$, if an ensemble of $M$ events is drawn?
\end{abstract}
\maketitle

\section{Introduction}
The probability $p(i)$ to draw an event $i$ from a set of $N$ possible
events is defined as the limes
\begin{equation}
p(i)=\lim\limits_{M\rightarrow\infty}\frac{M_i}{M}\,,~\mbox{with}~~
\sum\limits_{j=1}^N M_j = M
\end{equation}
where $f(i) \equiv M_i/M$ is the relative frequency to find the $i$-th
event $M_i$ times in a randomly chosen sample of $M$. According to the
law of large numbers the relative frequencies stochastically converge
to the corresponding probabilities, hence, for very large sample size
$M\gg N$ we can practically identify both values. For smaller sample
size, however, the distribution of relative frequencies may deviate
significantly from the probability distribution
(e.g.~\cite{Schmitt,PER,Herzel} and many others). Assume further the
equidistribution $p(i)=1/N$, then for very large sample size $M\gg N$
one expects that all or almost all of the $N$ possible events are
found in the sample and occur with approximately equal frequency,
whereas in the opposite case $M\ll N$ almost all events occur only
once or a few times in the sample, i.e. the latter frequency
distribution will deviate significantly from the former one. From this
simple argumentation one may conclude that the frequency distribution
which one expects depends sensitively on the sample size $M$. This
type of finite size effects is of major relevance for statistical
analysis of DNA and other biosequences, e.g.
\cite{Allegrini,Bernaola,Holste,Peng}. In this article we want to
calculate the expected frequency distribution which one finds in
dependence on the sample size $M$.

If we draw a frequency distribution of $N$ different
events there are $N!$ possibilities to arrange the events along the
abscissa. An arrangement which leads to a decaying function for the
frequencies or probabilities we call a Zipf order and the
corresponding distribution a Zipf ordered distribution. To find a Zipf
ordered frequency distribution which we have to expect if $M$ events
from an equidistribution are drawn we have to determine in dependence
on $M$ how many events (on average) are not drawn, i.e.~are drawn zero times,
how many are drawn once, twice etc.

In the next section we derive the expectation value for the number
$\langle K_i\rangle$ of those events which ocurr $i$ times in the
sample. The analytic expression is then used to infer the unknown
number $N$ of total events and to compare the theoretically expected
Zipf ordered frequency distribution with the measured one. The results
are useful in connection with entropy estimates computed from finite
samples.

\section{The number $K_i$ of different events each occurring
  exactly $i$ times and its expectation value $\langle K_i\rangle$}
\begin{figure}
\centerline{\psfig{figure=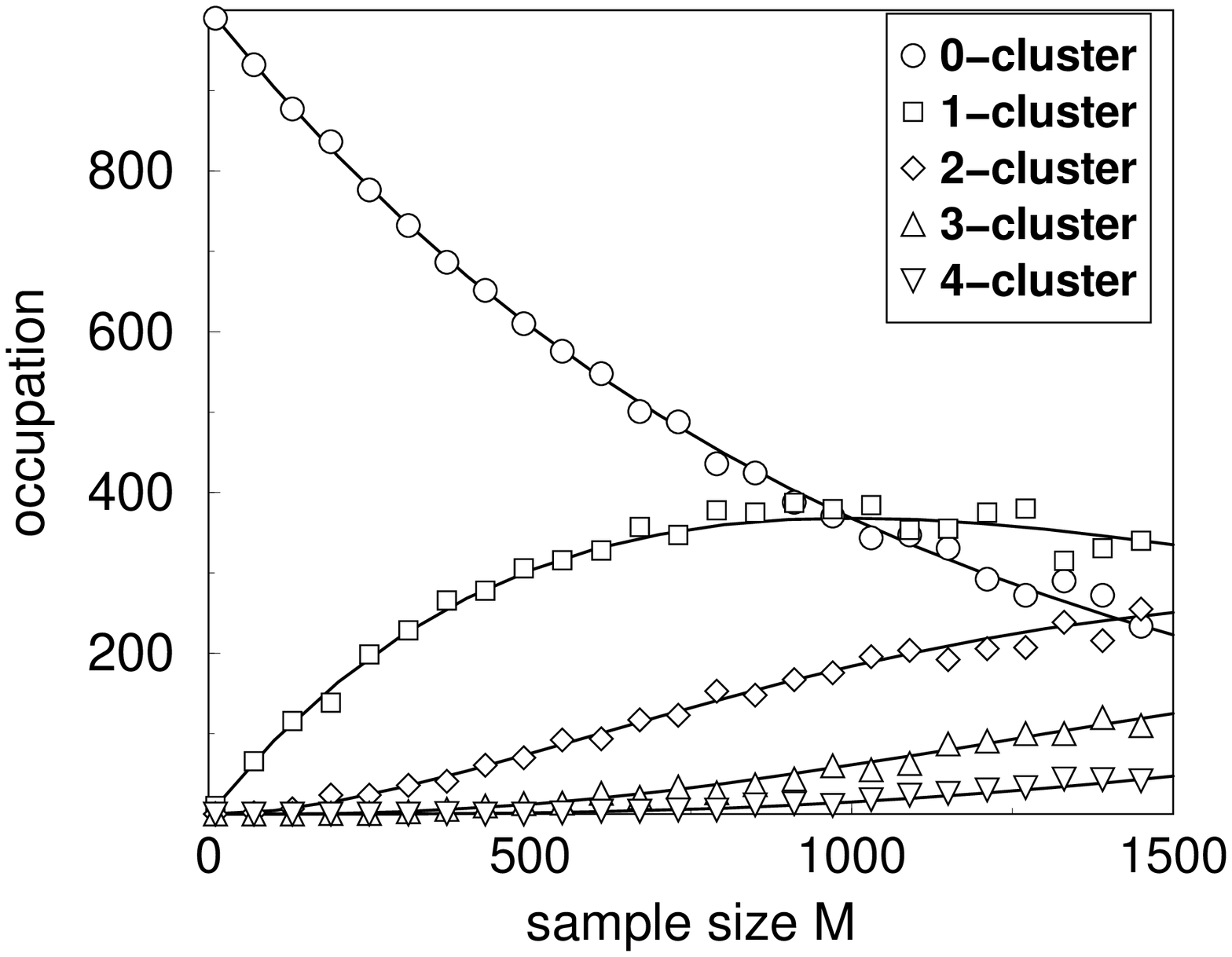,width=7cm}}
\centerline{\psfig{figure=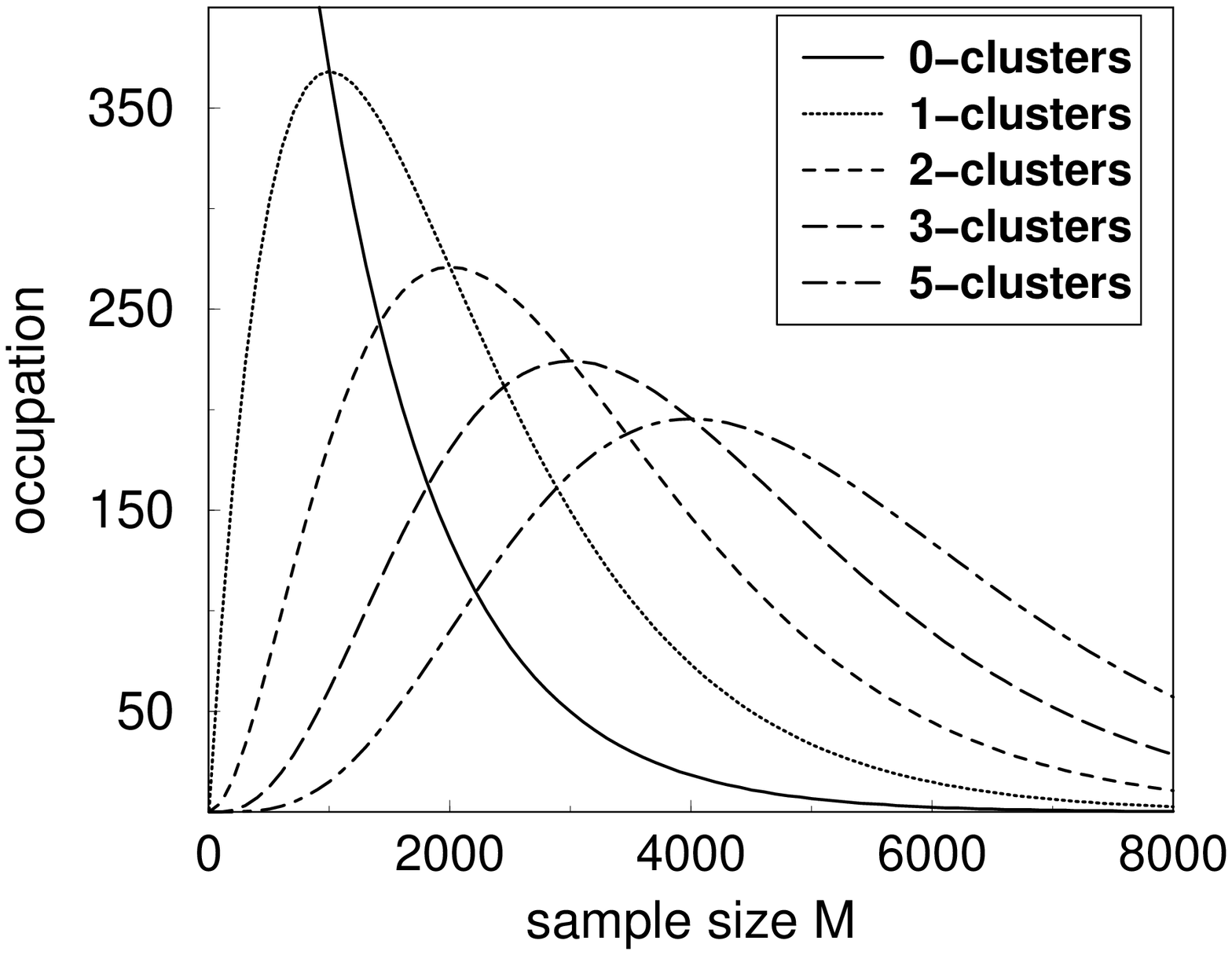,width=7cm}}
\caption{Expectation values for clusters of different sizes 
over the sample size $M$ taken from a set of $N=1000$ equidistributed
different events. The lines show the theoretical result calculated
from Eq. (\ref{eq:Mom}).  The symbols show the cluster distribution
found from a numerical simulation where $M$ random integers have been
drawn from the interval $[1,1000]$. The lower figure shows the same
data for a larger range of sample sizes.}
\label{fig:Mom}
\end{figure}

  The result of $M$ subsequent drawings from a set of $N$ different
  equiprobable events can be identified with randomly placing $M$
  indistinguishable balls in $N$ indistinguishable urns, each having
  the same probability $N^{-1}$. Denoting the number of urns
  containing exactly $i$ balls by $k_i$, a possible outcome can be
  shortly described by the vector $(k_0,k_1,\ldots,k_M)$; this is what
  we call a cluster configuration. The number of empty urns is given
  by $k_0$ and, consequently, the number of occupied urns by
  $N-k_0$. Any admissible cluster configuration obeys the following
  two conditions
\begin{eqnarray}
\label{bc1} 
\sum\limits_{i=0}^{M} k_i = N&&~~~\mbox{(total number of urns)}\\
\sum\limits_{i=0}^{M} k_ii = M&&~~~\mbox{(total number of balls)}
\label{bc2}
\end{eqnarray}
We are interested in the stochastic variable $K_i$, denoting
the number of urns each filled with exactly $i$ balls, and its
expectation value $\langle K_i\rangle$. Introducing for each
$i=0,1,\ldots$ and each urn $j=1,\ldots,N$ its related indicator
$I_i(j)$ by the following definition
\begin{equation}
\label{indicator}
I_i(j) = \Bigg\{ 
\begin{array}{l@{\qquad}l}
1 & \mbox{if urn $j$ contains exactly $i$ balls}\\
0 & \mbox{else\qquad\qquad}\,,
\end{array}
\end{equation}
the random variable $K_i$ is related to the stochastic indicators by
$K_i= \sum_{j=1}^N I_i(j)$. Due to the additivity of the expectation
operator we find
\begin{equation}
\langle K_i\rangle = 
\left<\sum_{j=1}^N I_i(j)\right> = 
\sum_{j=1}^N\langle I_i(j)\rangle = 
N \langle I_i(j)\rangle\,,
\end{equation}
where we have used that all $\langle I_i(j)\rangle$ are identical.
The probability to find exactly $i$ balls in any of the urns (here labeled
$j$) and the remaining balls distributed arbitrarily among the 
remaining $N-1$ urns is the binomial distribution, hence,
\begin{equation}
 \langle K_i\rangle= 
 N\left(M\atop i\right) \frac{1}{N^i}
 \left(1-\frac{1}{N}\right)^{(M-i)}\,.
\label{eq:Mom}
\end{equation}
The expectation value $\langle K_i\rangle$ indicates how often events
in a sample of size $M$ occur exactly $i$ times on average. We call
these occupation numbers $i$-clusters. Obviously, for small $M\ll N$
nearly all of the $N$ possible events are 0-clusters, i.e., they do
not occur in our sample.  As $M$ increases the number of single
occupations increases as well.  For still growing $M$ the number of
multiple occupation becomes larger and, therefore, the number of
1-clusters decreases as more and more events occur multiple times in
the sample.  Figure \ref{fig:Mom} shows the occupation of the $N$
possible different events as a function of the sample size $M$. The
lines show the theoretical result Eq.  (\ref{eq:Mom}) and the symbols
in the left of Fig.~\ref{fig:Mom} show the clusters as they have been
found in sets of random numbers.

\begin{figure}
\centerline{\psfig{figure=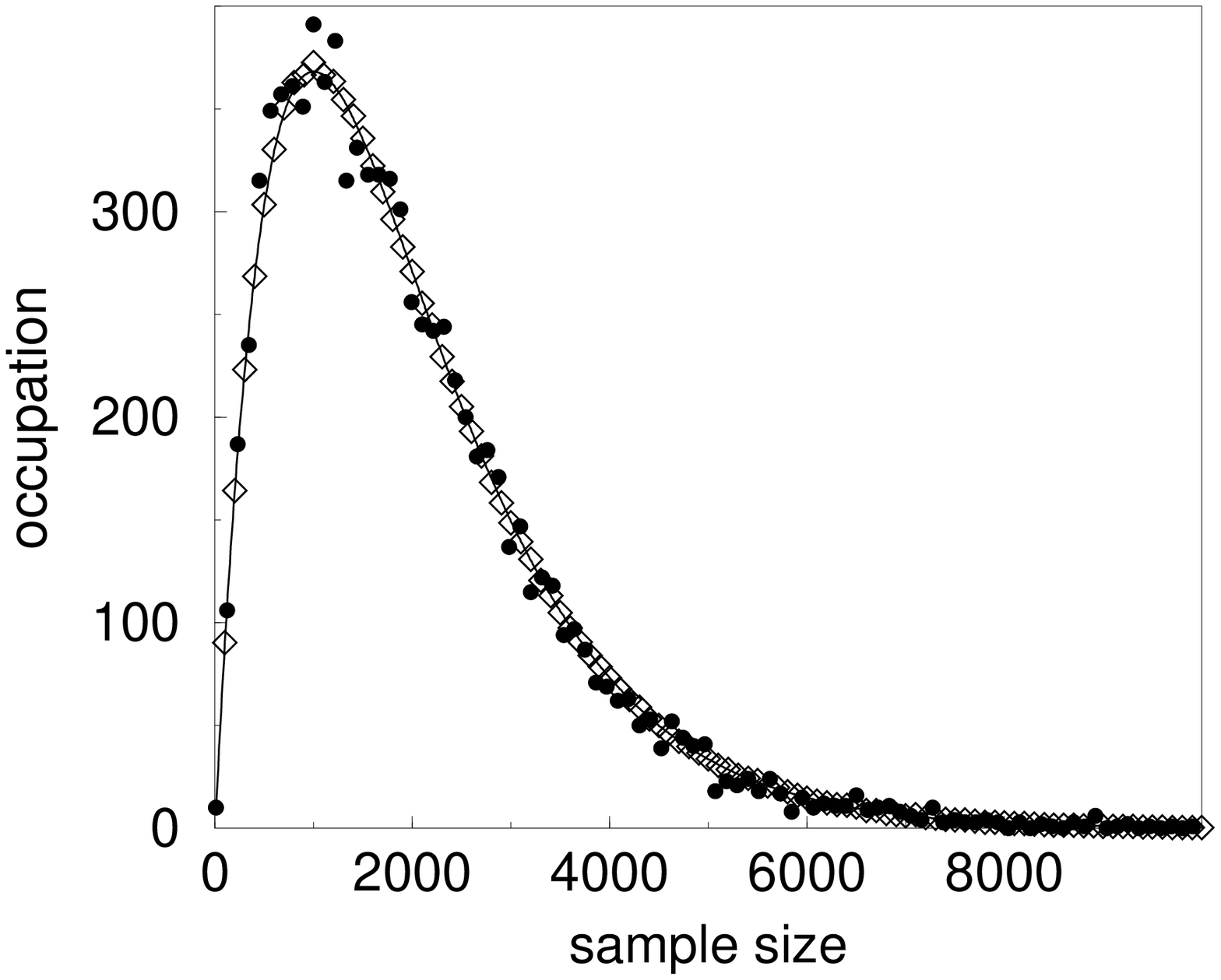,width=7cm}}
\centerline{\psfig{figure=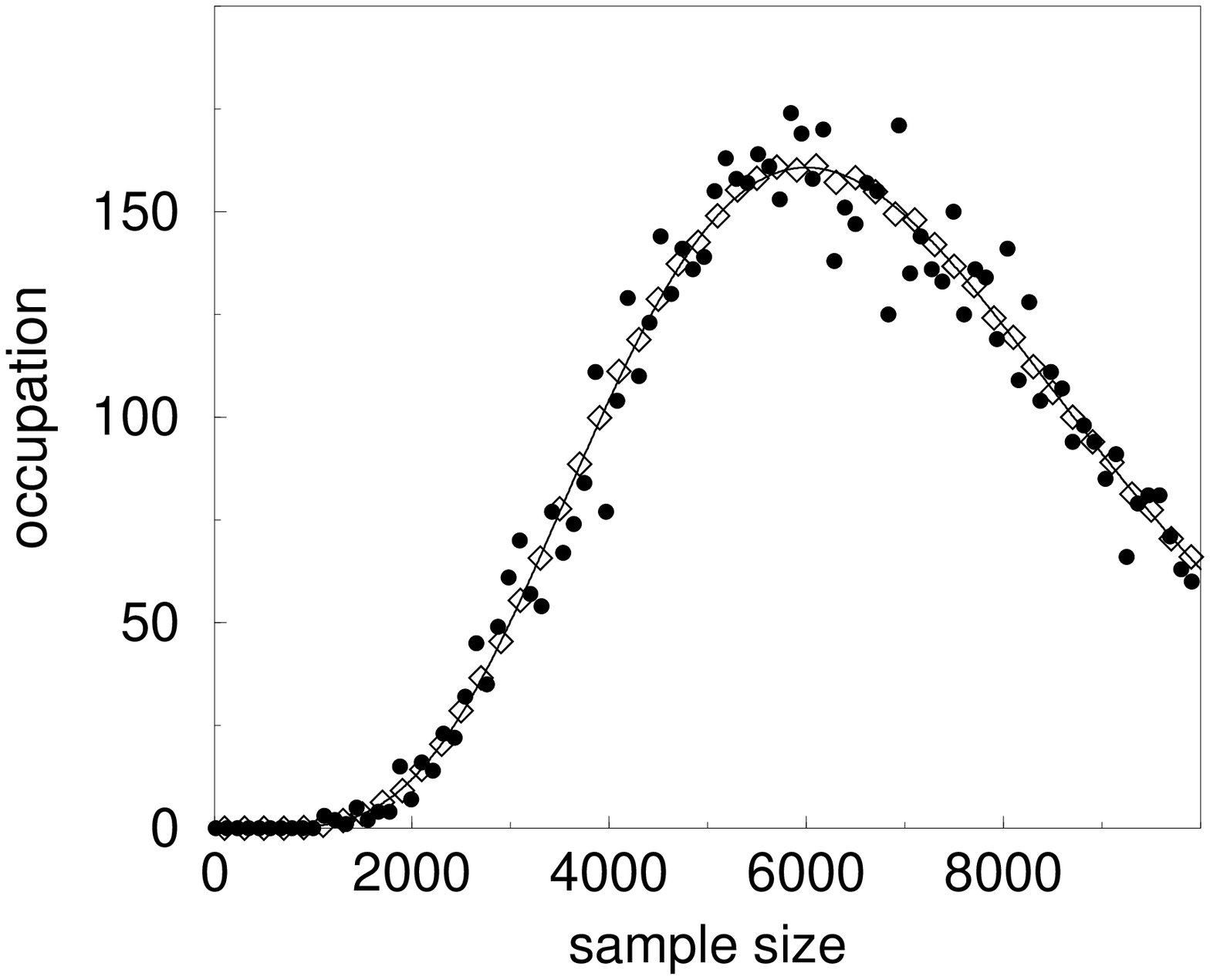,width=7cm}}
\caption{If samples of $M$ events are drawn independently, 
the resulting averaged cluster distribution converges to the
theoretical value Eq. (\ref{eq:Mom}). The filled dots show the
distribution of clusters of size 2 (top) and 6 (bottom) resulting from
a single drawing. The diamonds are averaged cluster distributions over
100 independent drawings of random numbers. The solid lines are the
theoretical values according to Eq. (\ref{eq:Mom}).}
\label{fig:Konv}
\end{figure}

If we draw only once a sample of size $M$ from a set of $N$ possible
events and calculate the occupation numbers (the cluster frequencies)
the cluster distribution itself is fluctuating (Fig.~\ref{fig:Konv},
filled dots). Averaging the cluster distribution over a number of
independent selections each of size $M$ the cluster distribution
converges to the theoretical curve predicted by
Eq. (\ref{eq:Mom}). Figure \ref{fig:Konv} shows the cluster
distribution as found from a random sample of size $M$ out of $N=1000$
allowed events together with cluster distributions averaged over 100
independent drawings.

\section{Zipf ordered frequency distribution}
\begin{figure}
\centerline{\psfig{figure=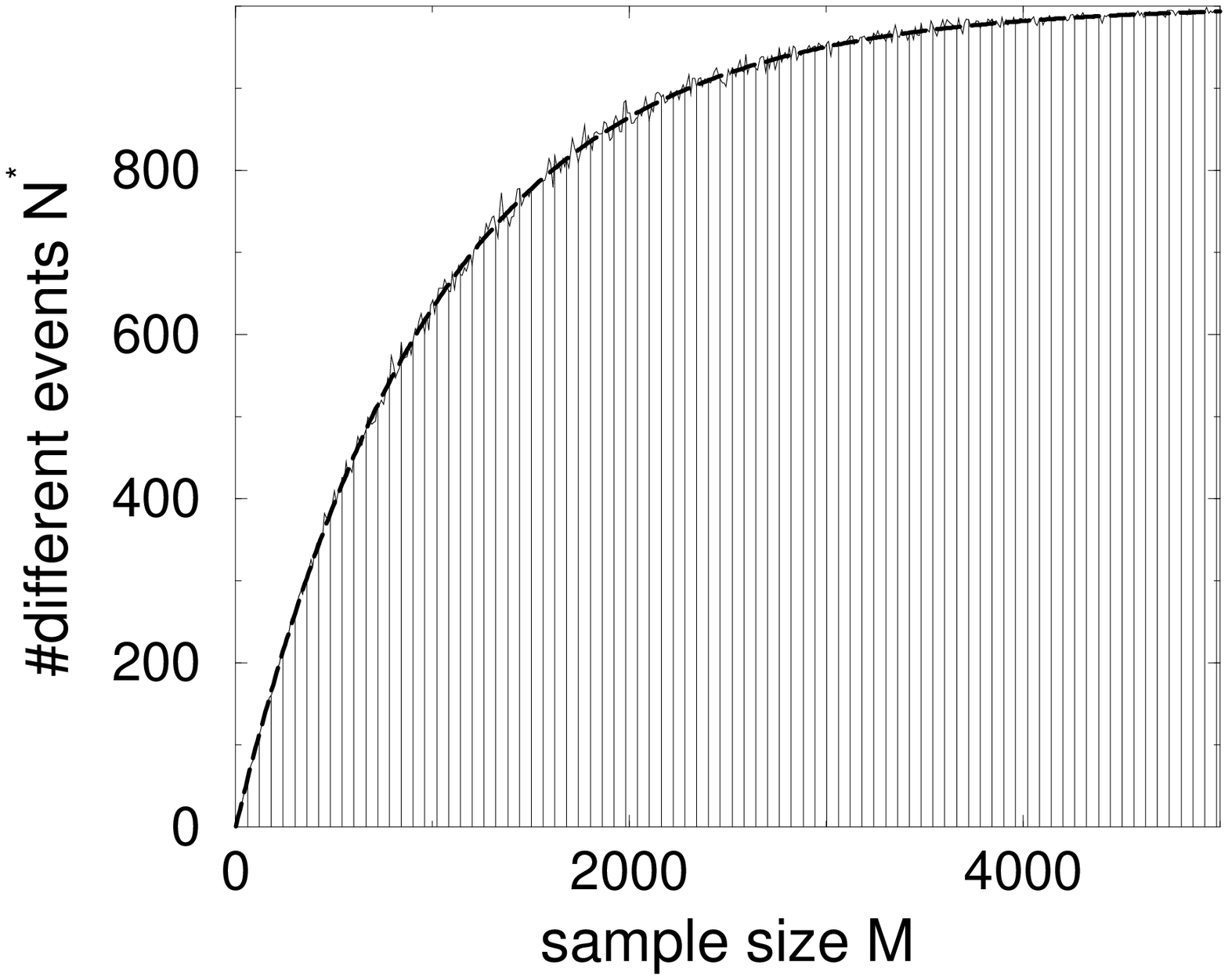,width=7cm}}
\centerline{\psfig{figure=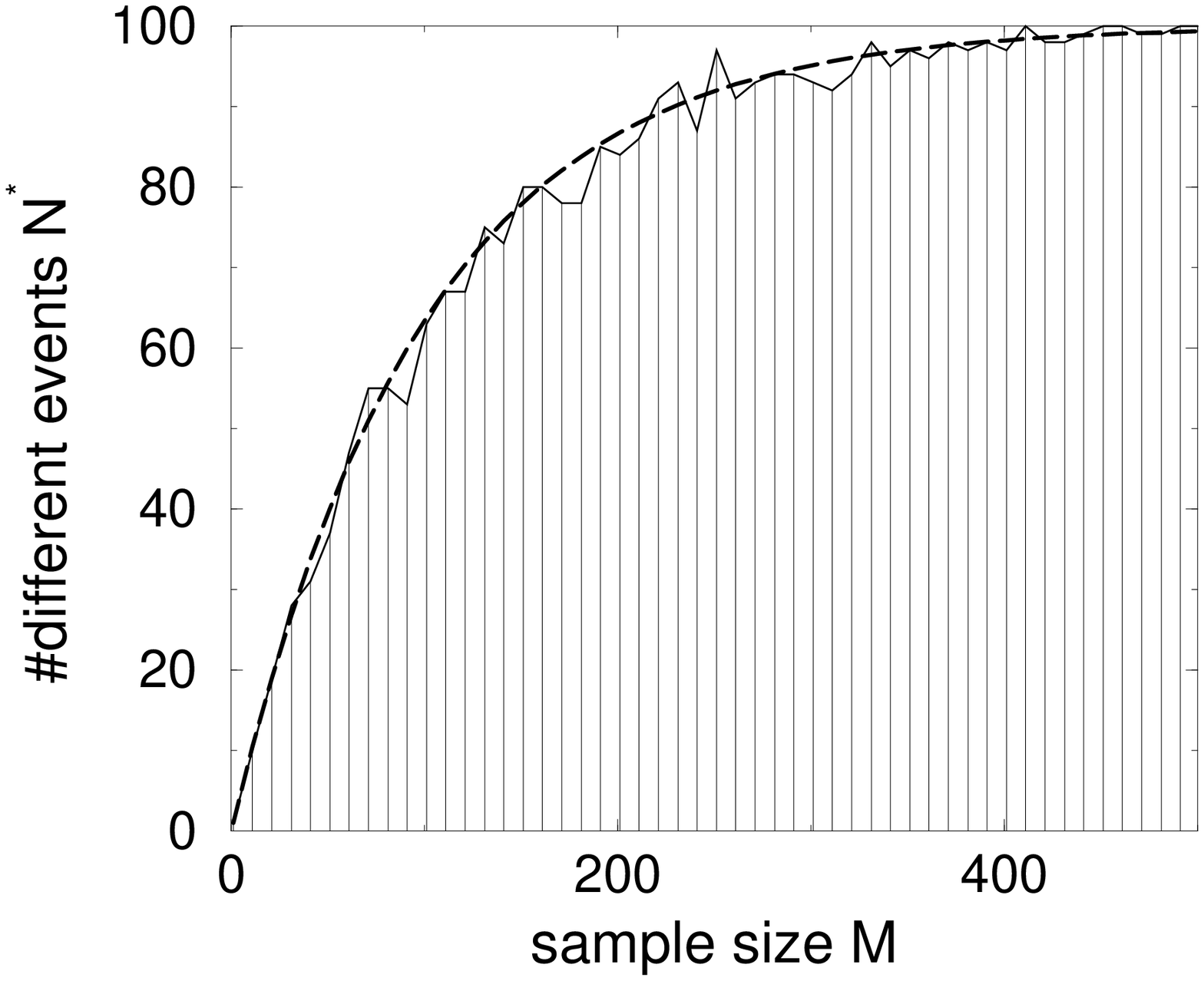,width=7cm}}
\caption{Number of different events $N^*$ in a sample of size $M$ 
drawn from a equidistribution $p_i=1/N$. The dashed line shows the
analytical result Eq. (\ref{eq:Nstar}), the impulses show the results
of a computer simulation. top: $N=1000$, bottom: $N=100$.}
\label{fig:Nstar}
\end{figure}

Equation (\ref{eq:Mom}) allows to determine the expectation value of
the number of different events $N^*$ in dependence on the total number
of drawn events $M$, which is simply related to the expected
probability to find a cluster of size zero:
\begin{equation}
N^*=N-\langle K_0 \rangle\,,
\end{equation}
from which we compute
\begin{eqnarray}
\label{eq:Nstar}
\frac{N^*}{N}
&=&
1-\left(1-\frac{1}{N}\right)^M=
1-\exp\left[M\log\left(1-\frac{1}{N}\right)\right] \nonumber\\
&=&
1-\exp\left(-\frac{M}{N}\right)\cdot\exp\left[-{\cal O}
\left(\frac{M}{N^2}\right)\right]
\label{eq:Nstart}
\end{eqnarray}
From Eq.~(\ref{eq:Nstart}) we see that for all $M\ll N^2$ Eq.
(\ref{eq:Nstar}) can be approximated to very good accuracy by
\begin{equation}
\frac{N^*_a}{N}\approx 1-\exp \left(-\frac{M}{N}\right)\,,
\label{eq:Nstarapprox}
\end{equation}
a result which has been found before by computer simulations
\cite{rosa}. The maximal {\em absolute} deviation is
$N^*-N^*_a=1/e\approx 0.37$ (for $N=M=1$) which falls rapidly to
$1/(2e)\approx 0.18$ as $N$ goes to infinity. In Fig.~\ref{fig:Nstar}
the analytical result (\ref{eq:Nstar}) is compared with a
simulation. The impulses in the figure show the results of a single
realization. If we average the numerical results over several runs the
numerical curve falls together with the analytical one.

For the wide range of practical interest, $5/8 \le N_a^*/M < 1$, from
Eq.~(\ref{eq:Nstarapprox}) we may approximate the entropy of the
distribution if we know the number of different events $N^*$ contained
in a sample of size $M$:
\begin{equation}
S\approx {\rm ld}\, \left( \frac{\left(3 M+\sqrt{3 M\left(8
N^*-5M\right)}\right)M}{12\left(M-N^*\right)}\right)\,.
\end{equation}

We want to compile the results from the previous section to find the
desired Zipf ordered frequency distribution. Equation (\ref{eq:Mom})
tells how many, in average, events do not occur in the sample (drawn
zero times), how many are drawn once, twice, etc. This yields directly
the Zipf ordered frequency distribution
\begin{equation}
f(i)=\left\{\begin{tabular}{lll}
0 &for &$N\ge i > N-\left<K_0\right>$\\
1 &for &$N-\left<K_0\right> \ge i 
  >  N-\left<K_0\right> -\left<K_1\right>$ \\
  &\dots&\\
$j$ &for &$N-\sum\limits_{k=0}^{j-1}
  \left<K_k\right> \ge i > N-\sum\limits_{k=0}^{j}\left<K_k\right>$\,.
\end{tabular}
\right.
\label{eq:Hauf}
\end{equation}
Using Stirling's formula to expand the expressions in
Eq. (\ref{eq:Mom}) the analytical result Eq. (\ref{eq:Hauf}) can be
written easily in elementary functions.

\begin{figure}
\centerline{\psfig{figure=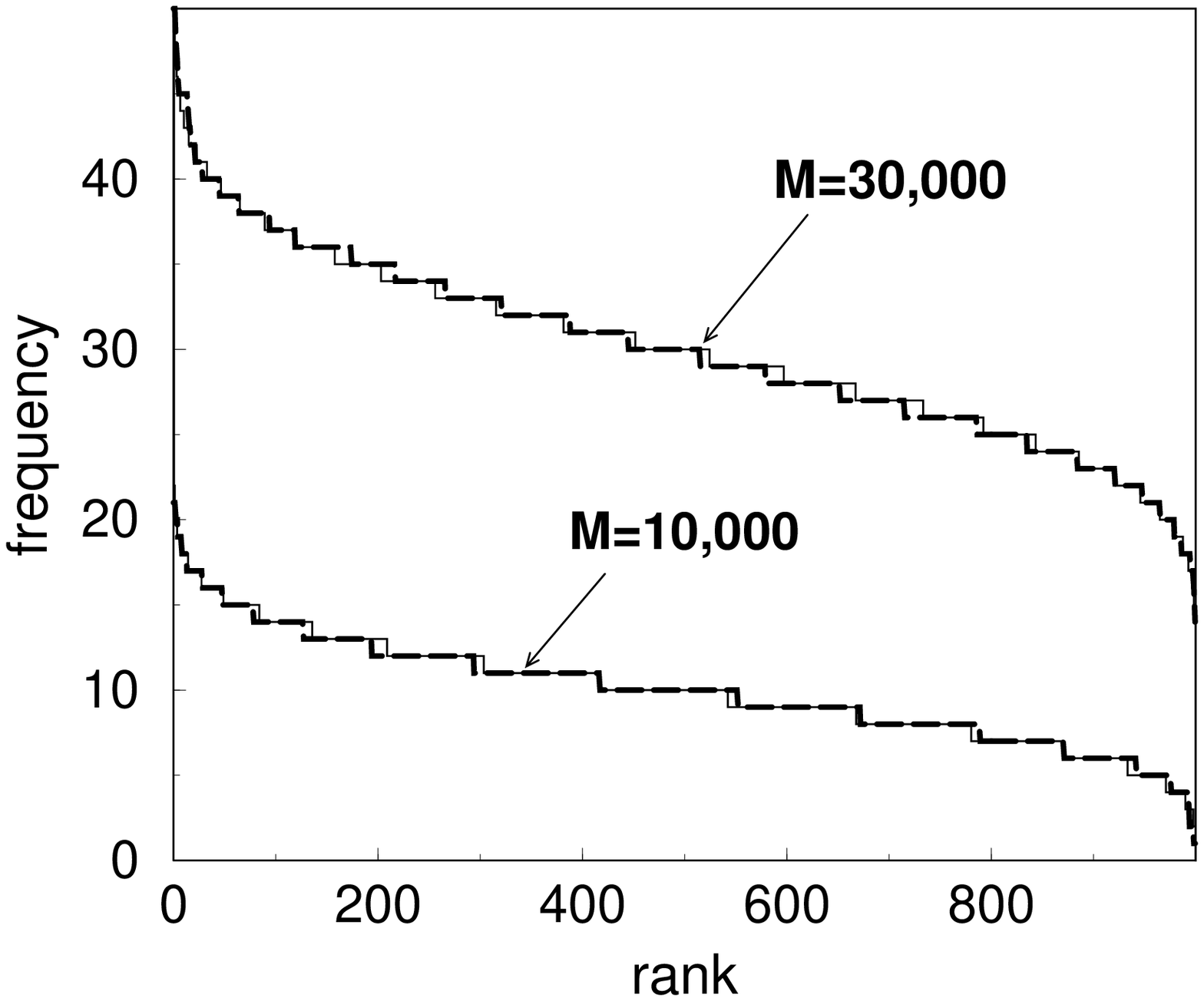,height=6.9cm}}
\centerline{\psfig{figure=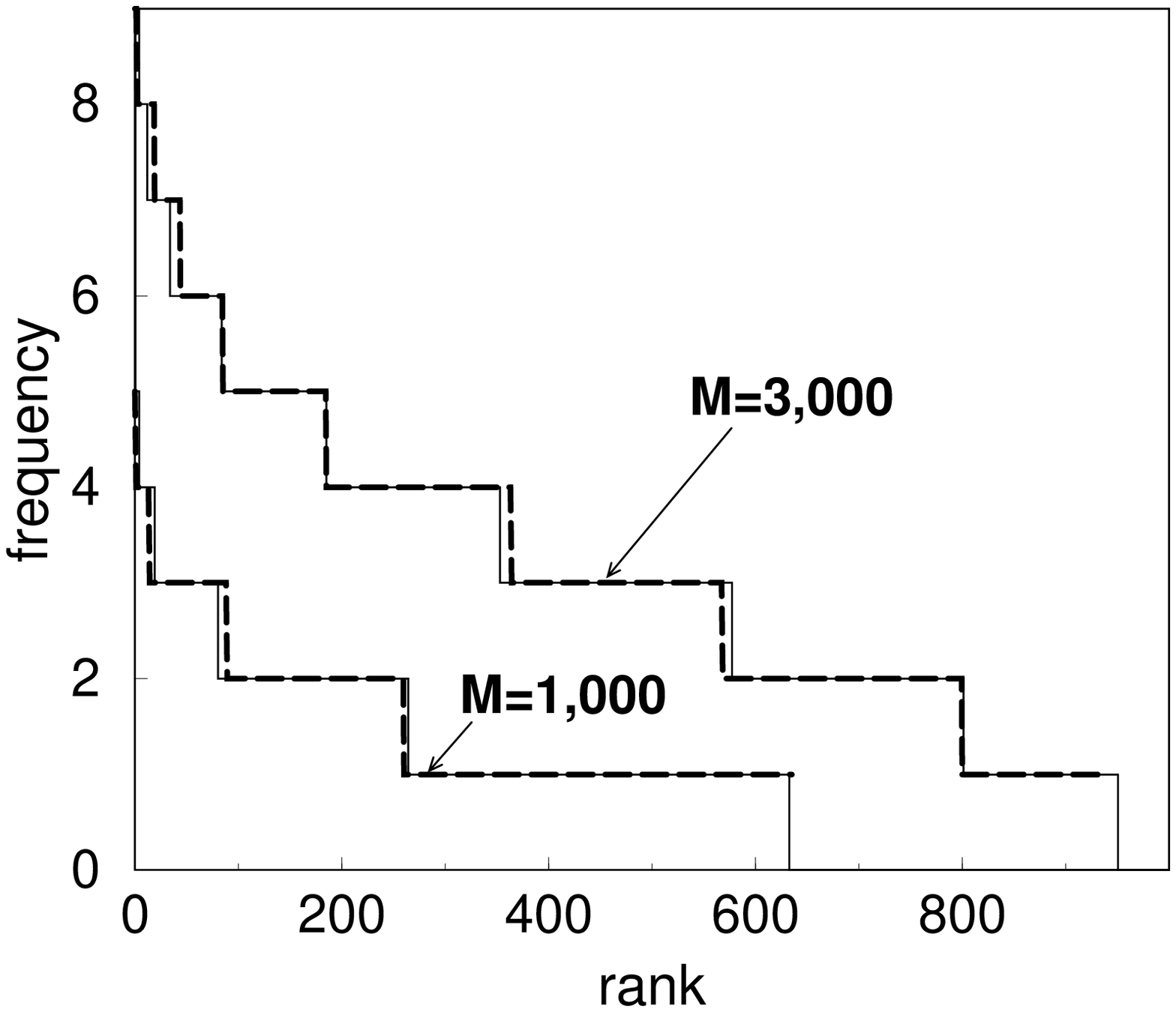,height=6.9cm}}
\caption{Zipf ordered frequency distributions calculated from samples 
of size $M$ from a equiprobability distribution of $N=1000$ different
random numbers (dashed lines). The theoretical curve due to
Eq. (\ref{eq:Hauf}) is drawn with solid lines.}
\label{fig:Zipf}
\end{figure}

Figure \ref{fig:Zipf} shows Zipf ordered frequency distributions
calculated from a sample of random numbers (dashed lines) together
with the theoretical distributions due to Eq. (\ref{eq:Hauf}) (solid
lines). The combinatorial theory derived in the previous section
predicts the Zipf ordered frequency distribution which results from an
equidistribution with good accuracy.

\section{Discussion}
Based on combinatorial considerations we calculated the expectation
values to find $k_i$ events {\em exactly} $i$ times in a sample of
size $M$ which have been drawn from an equidistribution. These cluster
probabilities allow to estimate the total number of events $N$,
given the number of {\em different} events found in a sample of size
$M$ is known. Moreover, the full Zipf ordered frequency distribution
could be constructed. By numerical simulations it has been
demonstrated that the analytically derived values coincide with
``experimental'' results, i.e., with cluster distributions and Zipf
ordered frequency distributions originating from finite samples of
random numbers.
\medskip
\medskip

\begin{acknowledgments}
Discussions with Werner Ebeling and Rosa Ram\'{\i}rez
are acknowledged.
\end{acknowledgments}

%\bibliography{Distr}

\end{document}